\documentclass[twocolumn,english,prb,superscriptaddress]{revtex4-1}
\usepackage[T1]{fontenc}
\usepackage[latin9]{inputenc}
\usepackage{dsfont,braket}
\usepackage{color}
\usepackage{amsmath}
\usepackage{amssymb}
\usepackage{graphicx}
\usepackage{esint}
\usepackage{ulem}   % to strike things out
\makeatletter
\@ifundefined{textcolor}{}
{%
 \definecolor{BLACK}{gray}{0}
 \definecolor{WHITE}{gray}{1}
 \definecolor{RED}{rgb}{1,0,0}
 \definecolor{GREEN}{rgb}{0,1,0}
 \definecolor{BLUE}{rgb}{0,0,1}
 \definecolor{CYAN}{cmyk}{1,0,0,0}
 \definecolor{MAGENTA}{cmyk}{0,1,0,0}
 \definecolor{YELLOW}{cmyk}{0,0,1,0}
}

\@ifundefined{definecolor}
 {\usepackage{color}}{}
\@ifundefined{definecolor}{\usepackage{color}}{}
\@ifundefined{definecolor}{\usepackage{color}}{}
\@ifundefined{definecolor}{\usepackage{color}}{}

\usepackage{babel}

\makeatother

\usepackage{babel}
\begin{document}

\title{Characterization and stability of a fermionic $\nu=1/3$ fractional Chern insulator}

\newcommand{\mike}[1]{ { \color{red} \footnotesize (\textsf{MZ}) \textsf{\textsl{#1}} }}
\newcommand{\johannes}[1]{ { \color{purple} \footnotesize (\textsf{JM}) \textsf{\textsl{#1}} } }
\newcommand{\adolfo}[1]{ { \color{blue} \footnotesize (\textsf{AG}) \textsf{\textsl{#1}} } }
\newcommand{\frank}[1]{ { \color{green} \footnotesize (\textsf{FP}) \textsf{\textsl{#1}} }}

\author{Adolfo G. Grushin}
\affiliation{Max-Planck-Institut f\"ur Physik komplexer Systeme, 01187 Dresden, Germany}
\author{Johannes Motruk}
\affiliation{Max-Planck-Institut f\"ur Physik komplexer Systeme, 01187 Dresden, Germany}
\author{Michael P. Zaletel}
\affiliation{Department of Physics, University of California, Berkeley, California 94720, USA}
\author{Frank Pollmann}
\affiliation{Max-Planck-Institut f\"ur Physik komplexer Systeme, 01187 Dresden, Germany}

\date{\today}

\begin{abstract}
Using the infinite density matrix renormalization group method on an infinite cylinder geometry, 
we characterize the $1/3$ fractional Chern insulator state in the Haldane honeycomb lattice model
at $\nu=1/3$ filling of the lowest band and check its stability. 
We investigate the chiral and topological properties of this state through 
(i) its Hall conductivity, 
(ii) the topological entanglement entropy,
(iii) the $U(1)$ charge spectral flow of the many body entanglement spectrum, and
(iv) the charge of the anyons. 
In contrast to numerical methods restricted to small finite sizes, the infinite cylinder geometry allows us to access and characterize directly the metal to fractional Chern insulator transition. We find indications it is first order and no evidence of other competing phases. Since our approach does not rely on any band or subspace projection, we are able to prove the stability of the fractional state in the presence of interactions exceeding the band gap, as has been suggested in the literature. 
As a by-product we discuss the signatures of Chern insulators within this technique.
\end{abstract}

\maketitle

\section{Introduction.}
The goal of completely understanding and experimentally realizing 
fractional Chern insulator (FCI) states \cite{NSCM11,SGS11,RB11,TMW11,SGK11,WBR12,BL13,PRS13}, 
analogues of the fractional quantum Hall states \cite{L83} that do not require external magnetic fields, 
is by now a deep open problem in condensed matter physics~\cite{KR93,SAD05,PJ06,HSD07,MC09,SRM12,SS14}. There are several promising proposals for experimentally realizing FCI states, including the possibility of accessing this state in strained graphene \cite{GCS12} and in cold atomic \cite{CD13},
molecular \cite{YGL13} or periodically driven systems \cite{GGN14}. Nonetheless,
there are still fundamental open questions regarding the emergence and stability of this state, mainly due
to the fact that most of the current understanding of FCI states stems from the 
exact diagonalization (ED) of small clusters. Accessing larger cluster sizes and thereby
clear signatures of the putative FCI state within ED often 
involves a truncation of the large Hilbert space to focus only on a small, physically relevant
subspace. This is most commonly achieved by projecting the Hamiltonian to the partially filled band \cite{NSCM11,WBR12} 
or by tracing over some degrees of freedom \cite{KNC14}, assumed \emph{a priori} to be physically irrelevant. 
This has proven to be a successful strategy in revealing the hallmarks of the FCI state such as the topological ground state (GS) degeneracy 
on the torus and the fractional Hall conductivity \cite{KVD12,KD13}, among others. 
Intriguingly, there has also been numerical evidence supporting the idea that even when departing 
from the Landau level paradigm the FCI state is still robust. 
Within ED, it has been shown to survive  even when considering bands with a finite band dispersion \cite{LLB12,GNC12,ZSH13}, 
higher Chern numbers \cite{BQ12,LBF12,WHC12,YGS12,WRB13a,SRB13,WRB14}, and even allowing for the interaction
energy scale to exceed single particle band gaps \cite{HSD07,SGS11,SRM12,KNC14}. 
The question of whether these conventional and unconventional features survive in the thermodynamic 
limit, as well as the nature of phase transitions from the FCI into neighbouring phases is still open.\\ 
\begin{figure}
\includegraphics[width=8cm,page=1]{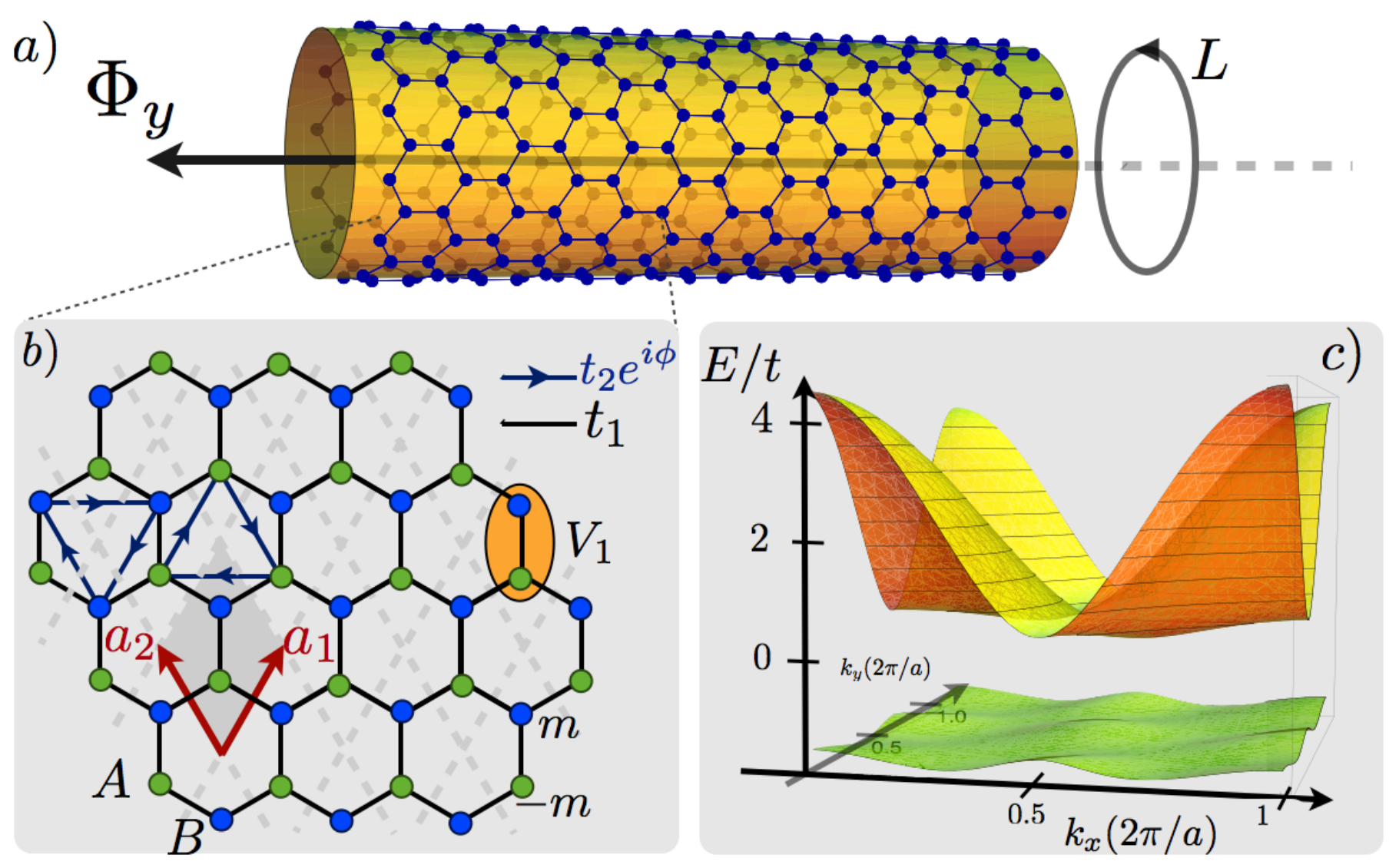}
\caption{\label{fig: CIcyl} 
(Color online) (a) Honeycomb lattice on an infinite cylinder with an $L$-site ($L=2L_{y}$) circumference implemented in iDMRG.
(b) Hopping conventions for the Haldane model \cite{H88} used in this work. (c) Band structure for the Haldane model with optimal band flatness ratio of the lower band ($\sim 1/7$), achieved with $\cos(\phi)=t_{1}/(4t_{2})=3\sqrt{3/43}$ ($\phi=0.65$) and $m=0$.}
\end{figure}
In this work, we study a fermionic FCI at filling $1/3$ on an infinite cylinder
using the infinite density matrix renormalization group (iDMRG) method~\cite{M08,W92,KZM13}.
We characterize the topological properties of the FCI phase and study its stability.  
Furthermore, we report numerical evidence
revealing the first order character of the phase transition from the FCI into a metallic state, occurring 
when interactions are of the order of the band width and find no evidence
of other competing orders.
Compared to ED, the iDMRG algorithm offers the advantage that we
can consider systems of size $L_x \times L_y$ unit cells with $L_x$ being in the thermodynamic limit
and $L_y$ larger than what is tractable by ED. Furthermore, we can 
conveniently  probe topological properties of the state and its quasiparticles, namely 
the topological entanglement entropy (EE) \cite{KP06,LW06,JWB12},
entanglement spectrum (ES)~\cite{LH08} and
the quasiparticle charges~\cite{HZQ13,ZMP13} encoded in the MPS representation of the wave function. 
Indeed, this method in its finite and infinite versions, has been recently shown to be remarkably successful in describing the topological
properties of fractional quantum Hall (FQH) states~\cite{ZMP13} and realizations 
of the $\mathbb{Z}_{2}$ quantum spin liquid~\cite{HSC14}, chiral spin liquid \cite{HSC14b} and 
bosonic FCI states \cite{JWB12,CV13,ZKB13}. The success of iDMRG for 2D systems 
relies on having low enough entanglement and short correlation lengths, a requirement which is commonly 
met by gapped systems.

\section{Model.} The  Haldane model for spinless fermions  with nearest-neighbor interactions  has the form~\cite{H88}
\begin{subequations}
\begin{eqnarray}\label{eq: Ham} 
H&=&H_{0}+H_{V},\\
H_{0}&=&-\sum\limits_{ij} t_{ij}c^{\dagger}_{i}c_{j}+\sum\limits_{i} m \left(n_{A,i}-n_{B,i}\right),\\
H_{V}&=&V_{1}\sum\limits_{\left\langle ij \right\rangle} n_{A,i}n_{B,j}.
\end{eqnarray}
\end{subequations}
The non interacting part $H_{0}$, is defined on the honeycomb lattice shown in Fig.~\ref{fig: CIcyl}(b) on the infinite cylinder of circumference $L=2L_{y}$ in Fig.~\ref{fig: CIcyl}(a). Electrons are created by $c^{(\dagger)}_{i}=a^{(\dagger)}_{i},b^{(\dagger)}_{i}$ with $n_{c,i}=c^{\dagger}_{i}c_{i}$ in each of two interpenetrating triangular lattices $A$ and $B$ spanned by the lattice vectors $\mathbf{a_{1,2}}=a(3/2,\pm3/2)$, where $a$ is the lattice constant. Hopping is allowed to nearest and next-nearest neighbors (NN and NNN) sites with amplitudes $t_{\left\langle ij \right\rangle}=t_{1} \in \mathbb{R}$ and $t_{\left\langle\left\langle ij \right\rangle\right\rangle}=t_{2}e^{\pm i\phi} \in \mathbb{C}$ respectively, where $\pm\phi$ is the phase acquired by an electron hopping between atoms in the same sublattice with the sign given by the direction of the arrows in Fig.~\ref{fig: CIcyl}(b). The staggered chemical potential $m$ controls the phase transition from a CI with Chern number $C=1$ to a trivial insulator with $C=0$~\cite{H88}. The band structure for this Hamiltonian in the CI phase is shown in Fig.~\ref{fig: CIcyl}(c) for the parameters that optimize the flatness ratio for the lower band, given in the figure caption. The interaction term $H_{V}$ is a NN short-range repulsive interaction represented by a shaded oval in Fig.~\ref{fig: CIcyl}(b).\\

We obtain the matrix product state (MPS) representation of the GS variationally using the iDMRG method~\cite{M08,W92,KZM13}. 
We represent the Hamiltonian using a matrix product operator (MPO) representation~\cite{ZMP13,KZM13}. The iDMRG algorithm finds iteratively an efficient representation of the ground state using a \emph{Schmidt decomposition} of the system into two half infinite cylinders and truncates it at a given (bond) dimension $\chi$. The Schmidt decomposition at a  bond is given by $\left|\psi\right\rangle =\sum_{\alpha}\Lambda_{\alpha}\left|\alpha_{L}\right\rangle\otimes\left|\alpha_{R}\right\rangle$, where $\Lambda_{\alpha}$ and $\left|\alpha_{L/R}\right\rangle$ are the Schmidt coefficients and left and right Schmidt states respectively. 
Importantly, such decomposition gives straightforward access to the entanglement properties. Note that the Schmidt values are directly related to the eigenvalues of the reduced density matrix, i.e,  $\rho^{L/R}_{\alpha} = \Lambda_{\alpha}^2$ and the Schmidt states are the corresponding eigenstates.
The von-Neumann entanglement entropy $S_{\mathrm{VN}}$ and the entanglement spectrum $\left\lbrace\varepsilon_{\alpha}\right\rbrace$  \cite{LH08} can be obtained by means of the relations $S_{\mathrm{VN}}=-\sum_{\alpha}\Lambda^{2}_{\alpha}\log\left(\Lambda_{\alpha}^2\right)$ and  $\Lambda_{\alpha}^{2}=e^{-\varepsilon_{\alpha}}$, respectively. 
Due to its faster numerical convergence, we will also find it useful to study the infinite Renyi entropy defined by $S_{\infty}=-\log\left(\mathrm{max}\left[\Lambda^2_{\alpha}\right]\right)$. Before discussing our results, let us note that, as in the FQH case \cite{ZMP13}, we take advantage of the $U(1)$ charge conservation symmetry of the Hamiltonian, which is also a symmetry of $\rho^{L/R}$. Thus the entanglement spectrum can be resolved further into distinct $U(1)$ charge sectors $Q^{L}_{\alpha}\in \mathbb{Z}$ where $Q^{L}_{\alpha}$ label the $U(1)$ charges of the left Schmidt states.

\section{Hall conductivity and spectral flow of the CI and FCI states.} 

The hallmark of the CI (FCI) state is the integer (fractional) quantization of the Hall conductivity $\sigma_{H}$.
This quantity can be written as an average of the Berry curvature over boundary conditions \cite{NTDW85} 
\begin{eqnarray}\label{eq:hall_bcond}
	\sigma_{H} &= \frac{e^2}{2\pi h} \int d\Phi_x d\Phi_y \, \nabla \times \mathcal{A}(\Phi_x, \Phi_y),
\end{eqnarray}
where $\Phi_x$ $(\Phi_{y})$ is the phase an electron acquires when a flux threads the cylinder in the $x$ ($y$) direction
and $\mathbf{\mathcal{A}}(\boldsymbol{\Phi}) = -i\Braket{ \Psi_0^{\boldsymbol{\Phi}}| \partial_{\boldsymbol{\Phi}} | \Psi_0^{\boldsymbol{\Phi}} }$ is the Berry connection computed from the GS wave function on the torus
$\left | \Psi_0^{\mathbf{\Phi}}\right\rangle$ for a flux $\boldsymbol{\Phi}=(\Phi_x,\Phi_y)$. In principle, calculating $\sigma_{H}$ involves repeatedly 
solving for the GS in a 2D discrete flux grid. 
However, iDMRG offers a great  simplification~\cite{ZMP14}. Fixing the flux through the cylinder $\Phi_y$, the $U(1)$ Berry phase for
one quantum of flux insertion in $\Phi_x $ is computed from the entanglement spectrum as
\begin{equation}
	e^{i \gamma(\Phi_y)} = \exp\big[ 2 \pi i\sum_\alpha \Lambda^2_{\alpha}(\Phi_y) Q^{L}_{\alpha}(\Phi_y) \big].
\label{eq:hall_gamma}
\end{equation}
Then, computing \eqref{eq:hall_bcond} reduces  to $\sigma_{H}=\frac{e^2}{2\pi h}\int_{0}^{2\pi}d\Phi_{y}\partial_{\Phi_{y}}\gamma(\Phi_y)= \frac{e^2}{h} \gamma(\Phi_{y})|^{2\pi}_{0}$ \cite{ZMP14}. The flux $\Phi_{y}$ threading through the cylinder is implemented in the MPO Hamiltonian by twisting the boundary conditions such that the electrons pick up a phase $e^{i\Phi_{y}}$ when circling around the cylinder. Previous ED calculations~\cite{KVD12} have obtained $\sigma_{H}$ either directly from Eq.~\eqref{eq:hall_bcond}, requiring a summation over excited states as well as an averaging over topologically distinct GS, or by computing the vorticity of the integrand of (2) directly through knowledge of the ground-state for different twist angles~\cite{HSD07}. Within the iDMRG framework $\sigma_{H}$ is computed in practice solely from the GS wave function at $\Phi_{y}=0,2\pi$.
Note that from the definition of the reduced density matrix, the exponent in Eq.~\eqref{eq:hall_gamma} is actually $2\pi i\left\langle Q^{L}(\Phi_{y})\right\rangle$, where $\left\langle Q^{L} (\Phi_{y})\right\rangle$ is the charge polarization modulo $1$ across a cut at fixed $x$ when flux $\Phi_y$ threads the cylinder.  As we adiabatically insert a flux $\Phi_{y}$, the latter quantity counts the number of charges crossing the entanglement cut~\cite{L83}.\\
In Fig.~\ref{fig: CI}(a) we show $\left\langle Q^{L}(\Phi_{y})\right\rangle$ for both the non-interacting trivial and CI cases at half filling for a cylinder with $L=6$ \footnote{The benchmarking and success of the iDMRG algorithm in describing trivial and CI phases at half-filling will be reported elsewhere \cite{MGP14}.}. The trivial insulator with $C=0$ (top blue) has no charge pumping after one flux quantum and thus $\sigma_{H}=0$. On the other hand, the CI state with $C=1$ shows that, after the insertion of one flux quantum, there is exactly one unit charge pumped across the boundary. By computing \eqref{eq:hall_bcond} as described above, we find an accurate quantization to $\sigma_{H;\mathrm{iDMRG}}=e^2/h$. Alternatively, we can visualize the charge pumping through the many-body ES $\left\lbrace\varepsilon_{\alpha}\right\rbrace$ shown in Fig.~\ref{fig: CI}(b). As mentioned above, the ES can be decomposed into different $U(1)$ corresponding to charge quantum numbers $Q_{\alpha}^{L}$. For a CI state, the adiabatic flux insertion must shift the ES by one charge sector after a single flux quantum is inserted, signalling that a unit charge has been pumped from left to right (or viceversa)~\cite{AHB11}. We find exactly such spectral flow numerically, shown in Fig.~\ref{fig: CI}(b). The calculation of the Hall conductivity and the spectral flow for the CI state obtained from iDMRG represents our first novel result.

Having understood the signatures of the CI state in iDMRG, we now address the existence and stability of the FCI state in a $\nu=1/3$ filled band of Hamiltonian Eq.~\eqref{eq: Ham}. For concreteness we focus first on the model with optimized parameters, i.e., we choose the optimal bottom band flatness ratio depicted in Fig.~\ref{fig: CIcyl}(c) and first set $V_{1}=t_{1}$, which exceeds the band width of the lower band $W\approx0.31t_{1}$ but is smaller than the band gap $\Delta\approx1.7t_{1}$.
In Fig.~\ref{fig: CI}(c), we show $\left\langle Q^{L}(\Phi_{y})\right\rangle$ and the entanglement spectrum as a function of flux $\Phi_{y}$ for up to three flux periods for a cylinder with circumference $L=12$ and $\chi=500$. From this plot, it is apparent that the GS wave-function obtained by iDMRG pumps one electron through the cut only upon insertion of \emph{three} flux quanta. We find that the Hall conductivity is $\sigma_{H;\mathrm{iDMRG}}\approx 0.33e^2/h$, consistent with a $1/3$ Laughlin-FCI state. Furthermore, the zero flux ES structure is shifted by one $U(1)$ particle sector only upon inserting three flux quanta [see Fig.~\ref{fig: CI}(d)], which provides evidence for a fractional charge pumping per cycle. We have checked convergence by confirming that larger values of $\chi$ give identical results. We emphasize that these results provide direct evidence for the charge pumping and entanglement spectral flow of an FCI state, not previously addressed with exact diagonalization. \\
\section{Characterization of topological order in the FCI state.} 
The iDMRG gives  direct access to a topological characterization of the FCI state~\cite{CV13,ZMP13,HZQ13}. For a topologically ordered state the finite size scaling of both $S_{\mathrm{VN}}$ and $S_{\infty}$ satisfies the area law $S=cL-\gamma_{i}$, with a non-universal constant $c$ but the same topological contribution $\gamma_{i}=\log\sqrt{\sum_{j}d^2_{j}}-\log(d_{i})$ where $d_{i}$ are the quantum dimensions of the different quasiparticles labelled by $i$~\cite{LW06,KP06}.
\begin{figure}
\includegraphics[angle=0,scale=0.33,page=1]{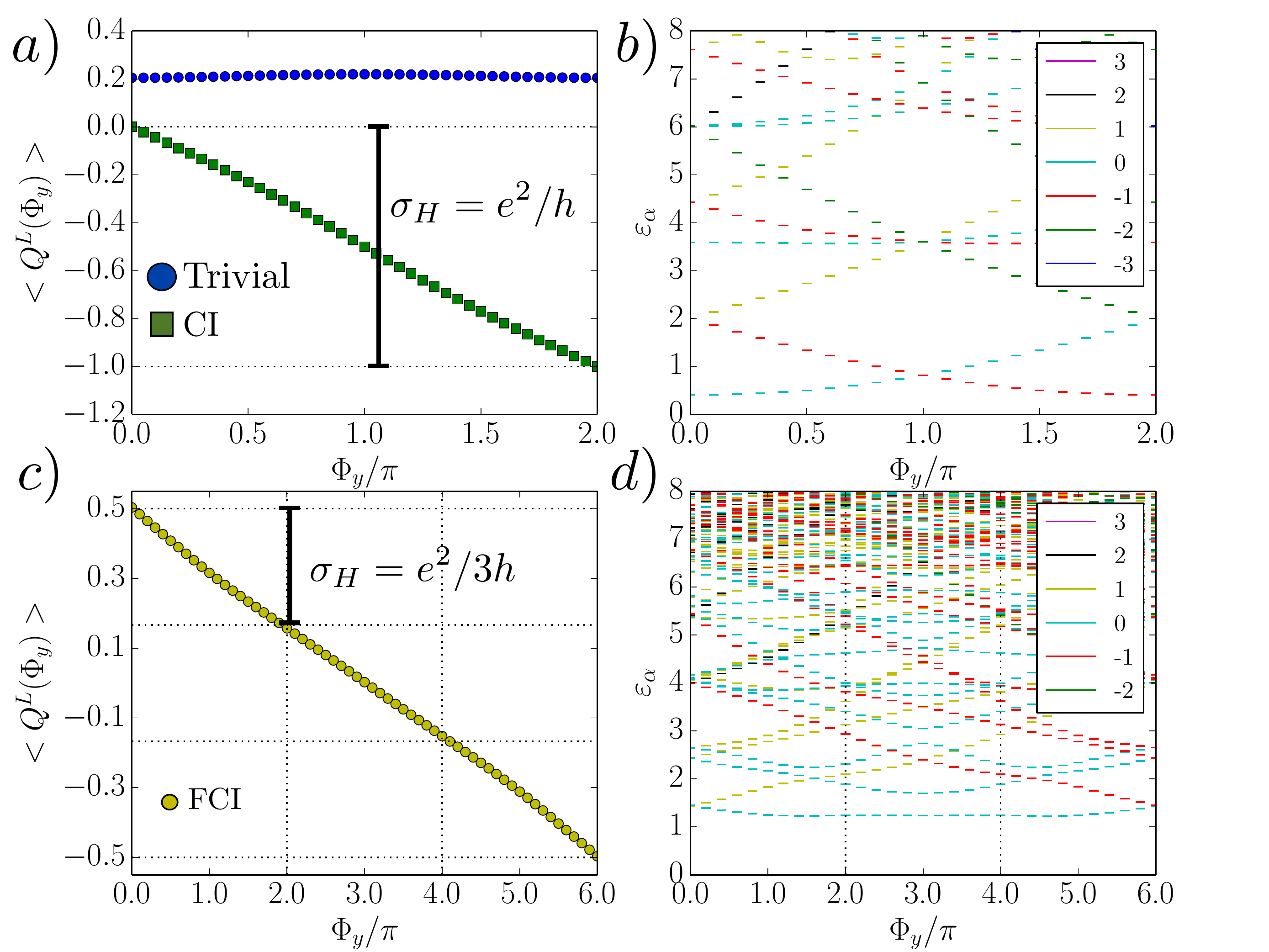}
\caption[Compact Routing Example]{\label{fig: CI} 
(Color online) (a) Charge pumping after one flux insertion for a half-filled trivial 
insulator on a cylinder with $L=6$ and $m\neq 0$, $t_{2}=0$ and $\chi=200$ (blue upper curve) and a CI with $m= 0$, $\phi\neq 0 , t_{2} \neq 0$ and $\chi=400$ (lower green curve). The bond dimension was chosen large enough to represent the groundstate for this size \cite{MGP14}.  (b) Entanglement spectrum evolution as a function of flux for the CI. The spectrum is shifted by a
unit charge after one flux quantum has been adiabatically inserted. Different charge sectors $Q^{L}$ are color coded. In (c) and (d) we show the charge pumping and ES for the $1/3$ filled lower band of the Haldane model after \emph{three} flux insertions for $L=12$ and $\chi=500$. In this case $\sigma_{H;\mathrm{iDMRG}}\sim 0.33e^2/h$ per flux quantum and the ES is shifted by a unit charge only after three flux quanta have been adiabatically inserted.}
\end{figure}
For the abelian $1/3$ Laughlin state, $d_{i}=1$ $\forall i=1,2,3$ and thus $\gamma=\log\sqrt{3}\sim 0.549$ \cite{LW06,KP06}. 
We show in Fig.~\ref{fig: FCI}(c)  that the extrapolation of the finite size scaling of $S$ 
is indeed in good agreement with this prediction resulting in $\gamma_{iDMRG}=0.50, 0.587$ for $S_{\mathrm{VN}}$ and $S_{\infty}$, respectively \footnote{For $S(L)$ in Fig.~\ref{fig: FCI} we include only those sizes that satisfy $L=6m$ with $m\in \mathbb{Z}$. Other sizes have a fractional charge per unit length on the cylinder in the Tao-Thouless limit and spontaneously break the symmetry into a charge density wave order which smears out as $L$ increases. The finite corrections for $S(L)$ when $L\neq 6m$ are thus expected to be more severe than for $L=6m$ and therefore we exclude them.}. 

Using the adiabatic flux insertion, we can probe all three topologically degenerate FCI GS (minimally entangled states) since they evolve continuously into one another every flux period. Thus, the three distinct topologically equivalent GS occur at $\Phi_{y}=2\pi m$ with $m=0,1,2$. Because of translational symmetry $T_{y}$ of the cylinder, we can assign momentum quantum numbers to Schmidt states~\cite{PT12,CV13} [see Fig.~\ref{fig: FCI} (b)]. We notice that the ES of all three GS have the characteristic counting of the edge conformal field theory (CFT), further confirming the topological nature of the state~\cite{KP06,LH08}. The access to all three GS enables us to extract the quasiparticle charge $q_{a}$ of the Laughlin state anyons~\cite{ZMP13,HZQ13}. Theoretically, the quasiparticle charges can be written as $q_{a}=(a+1/2)/3=\left\lbrace 1/2,1/6,-1/6 \right\rbrace$ mod 1~\cite{ZMP13}  where $a=\left\lbrace 1,0,-1 \right\rbrace$ labels the anyons and the $1/2$ factor takes into account the $\pi$ 
phase that is due to the periodic boundary conditions of the physical fermions. Numerically we can find the corresponding charges by noting that with each inserted flux quantum an anyon is pumped through a given cut. Thus we can directly read off from Fig.~\ref{fig: CI}(c) the quasiparticle charge from $q_{m}=\left\langle Q^{L}(\Phi_{y})\right\rangle|_{\Phi_{y}=2\pi m}=0.5,0.154,-0.154$ which are in close agreement with those expected theoretically (see Appendix \ref{app:1} for further details).
\section{Stability of the FCI state and the metal to FCI transition.} 
An important advantage of the iDMRG method is that we can obtain relatively large system sizes without subspace projection. 
This enables us to study the emergence and stability of the FCI state under decreasing/increasing $V_{1}$ in the presence of band mixing ~\cite{HSD07,SGS11,KNC14}. 
Importantly, we will also address the character of the transition between the FCI state and possible competing orders, an issue not previously discussed in the literature
to our knowledge but essential for benchmarking future experiments.

\begin{figure}
\includegraphics[angle=0,width=9.5cm,page=1]{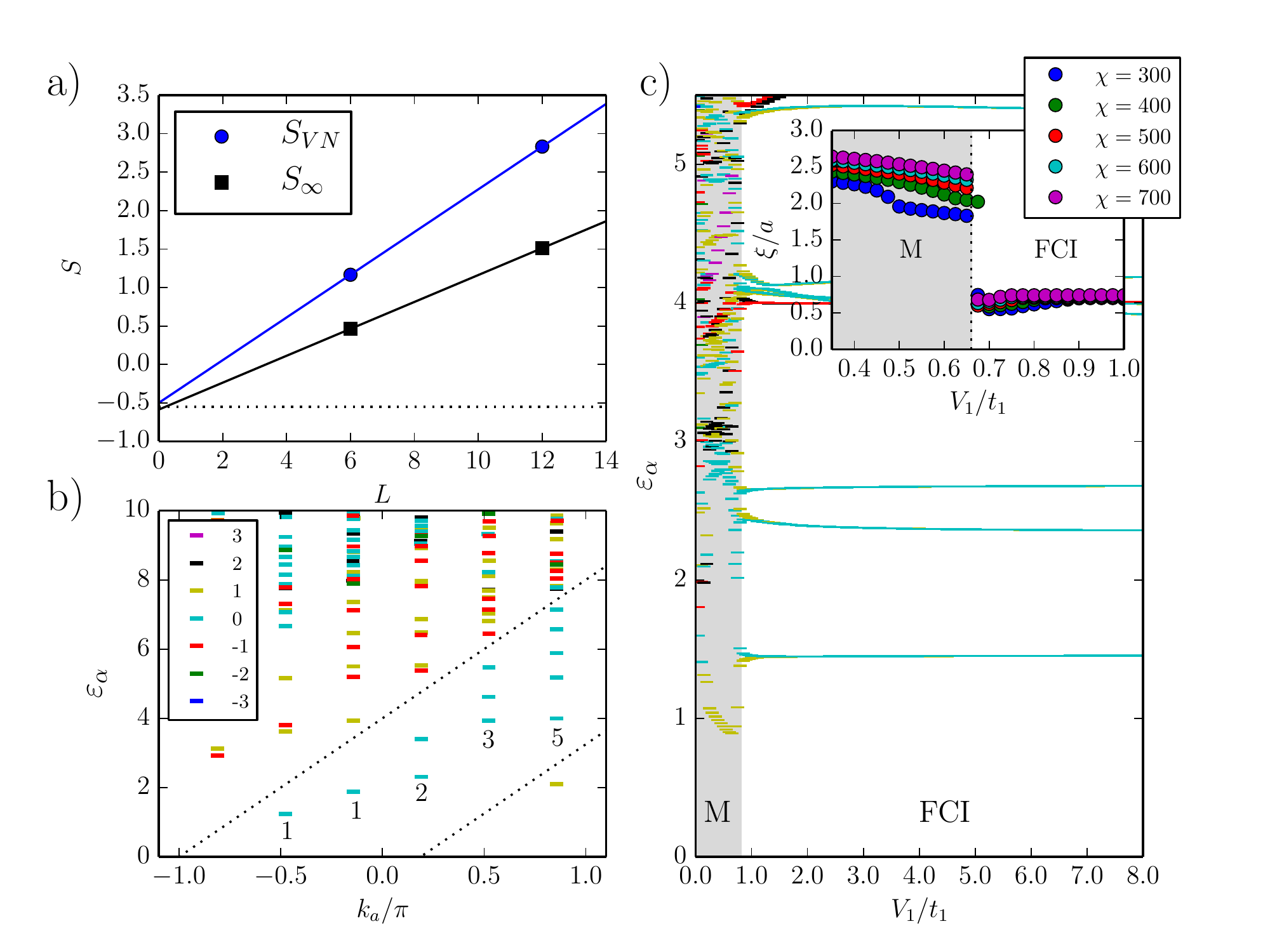}
\caption{\label{fig: FCI} 
(Color online) a) Finite size scaling of the von Neumann and infinite Renyi entropies as a function of cylinder circumference $L$ for $V_{1}=t_{1}$. They extrapolate to the values $\gamma_{\mathrm{VN}}=0.50$ and $\gamma_{\infty}=0.587$ respectively. The theoretically expected value $\gamma=\frac{1}{2}\log{3}$ is indicated by an horizontal dashed line. b) Momentum resolved entanglement spectrum of the GS at $\Phi_{y}=2\pi$ for $V_{1}=t_{1}$ showing the CFT edge theory counting $\left\lbrace 1,1,2,3,5, \cdots \right\rbrace $ for each $Q^{L}$ sector (labelled by different colors).  The dashed lines enclose the $Q^{L}=0$ sector counting for clarity. c) Entanglement spectrum $\varepsilon_{\alpha}$ as a function of $V_{1}/t_{1}$ with the same color coding as in b). The left shaded region is a metallic state (M) while the right unshaded region corresponds to the FCI state. Inset: Correlation length $\xi$ in units of $a$ as a function of $V_{1}/t_{1}$ for different $\chi$. The sharp discontinuity at $V_{1} \sim 2W$ signals a direct M-FCI phase transition.} 
\end{figure}

In Fig.~\ref{fig: FCI}(c) we show the ES as a function of $V_{1}$.
We find that there are two clearly distinct phases, represented by the shaded and unshaded regions. 
For $V_{1} \lesssim 2W$ we observe a phase with a dense ES and a correlation length $\xi$ which diverges as the MPS bond dimension $\chi$ is increased [see inset in Fig.~\ref{fig: FCI}(c)], which indicates the iDMRG is not converged in $\chi$.
This divergence is the expected behavior for a metallic phase,\cite{Tagliacozzo-2008,Pollmann-2009} because the logarithmically divergent entanglement of a metal requires an MPS with $\chi \to \infty$. 
In contrast, for $V_{1} \gtrsim 2 W$,  $\xi$ is well converged with $\chi$, indicating the state is gapped.
The contrasting behavior of $\xi$ is shown in inset in Fig.~\ref{fig: FCI}(c), which indicates a direct phase transition between a metal (shaded) and a FCI (unshaded).
Using Eq.~\eqref{eq:hall_gamma}, the entire FCI region $V_{1}\gtrsim 2W$ is found to have a robust Hall conductivity $\sigma_{H}=e^2/3h$. 
The existence of a metallic phase for $V_{1} \lesssim 2W$ is expected from a perturbative renormalization group perspective, \cite{S94} since the Fermi-liquid state is expected to be robust until $V_{1} \sim W$.
Note that $\xi$ remains constant in the entire FCI region, with no apparent divergence as the transition is approached; presumably this indicates the transition is first-order. For large $V_{1}$ we find no other singularities in the ES, implying that the FCI state is stable up to arbitrary interaction strength, consistent with ED results on small clusters~\cite{HSD07,SGS11,SRM12,KNC14}. Therefore, our numerical results rule out the appearance of any intermediate competing phases.

A related open question is the stability towards NNN interactions $V_{2}$. Preliminary tests indicate that for $V_{1}= t_{1}$ the state survives up to moderate $V_{2}\lesssim W$, although further study is needed to assess the complete phase diagram as a function of $V_{1,2}$. Finally, at $V_1= t_{1}$ we have also checked the FCI phase remains  for several values of the band parameters $t_2$, $\phi$ tuned away from the flatness-optimized point.
\section{Conclusions.}
We have characterized the fermionic Chern and fractional Chern insulator on the Haldane model within
iDMRG on an infinite cylinder. This numerical technique enabled us to fully characterize these states by directly accessing 
(i) chiral properties of the state, i.e. its Hall conductivity and entanglement spectrum spectral flow and 
(ii) topological properties of the FCI state, including its finite size entanglement scaling, the charge of the anyons and 
the counting of the entanglement spectrum. We have also exemplified how
to access different topologically degenerate ground states via flux insertion and shown that the state is robust against
band mixing or modifying the optimal parameters. We have presented for the first time the first order character of the metal to FCI transition from a sharp discontinuity in both the correlation length and the entanglement spectrum. Furthermore, our numerical results rule out the appearance of other competing phases, establishing a benchmark for future experiments.

\section{Acknowledgements:} We thank B. A. Bernevig for illuminating discussions, R. Mong for work on related problems
and N. Regnault for critical reading of the manuscript.

%\bibliography{FCI_iDMRG.bib}

\begin{thebibliography}{59}%
\makeatletter
\providecommand \@ifxundefined [1]{%
 \@ifx{#1\undefined}
}%
\providecommand \@ifnum [1]{%
 \ifnum #1\expandafter \@firstoftwo
 \else \expandafter \@secondoftwo
 \fi
}%
\providecommand \@ifx [1]{%
 \ifx #1\expandafter \@firstoftwo
 \else \expandafter \@secondoftwo
 \fi
}%
\providecommand \natexlab [1]{#1}%
\providecommand \enquote  [1]{``#1''}%
\providecommand \bibnamefont  [1]{#1}%
\providecommand \bibfnamefont [1]{#1}%
\providecommand \citenamefont [1]{#1}%
\providecommand \href@noop [0]{\@secondoftwo}%
\providecommand \href [0]{\begingroup \@sanitize@url \@href}%
\providecommand \@href[1]{\@@startlink{#1}\@@href}%
\providecommand \@@href[1]{\endgroup#1\@@endlink}%
\providecommand \@sanitize@url [0]{\catcode `\\12\catcode `\$12\catcode
  `\&12\catcode `\#12\catcode `\^12\catcode `\_12\catcode `\%12\relax}%
\providecommand \@@startlink[1]{}%
\providecommand \@@endlink[0]{}%
\providecommand \url  [0]{\begingroup\@sanitize@url \@url }%
\providecommand \@url [1]{\endgroup\@href {#1}{\urlprefix }}%
\providecommand \urlprefix  [0]{URL }%
\providecommand \Eprint [0]{\href }%
\providecommand \doibase [0]{http://dx.doi.org/}%
\providecommand \selectlanguage [0]{\@gobble}%
\providecommand \bibinfo  [0]{\@secondoftwo}%
\providecommand \bibfield  [0]{\@secondoftwo}%
\providecommand \translation [1]{[#1]}%
\providecommand \BibitemOpen [0]{}%
\providecommand \bibitemStop [0]{}%
\providecommand \bibitemNoStop [0]{.\EOS\space}%
\providecommand \EOS [0]{\spacefactor3000\relax}%
\providecommand \BibitemShut  [1]{\csname bibitem#1\endcsname}%
\let\auto@bib@innerbib\@empty
%</preamble>
\bibitem [{\citenamefont {Neupert}\ \emph {et~al.}(2011)\citenamefont
  {Neupert}, \citenamefont {Santos}, \citenamefont {Chamon},\ and\
  \citenamefont {Mudry}}]{NSCM11}%
  \BibitemOpen
  \bibfield  {author} {\bibinfo {author} {\bibfnamefont {T.}~\bibnamefont
  {Neupert}}, \bibinfo {author} {\bibfnamefont {L.}~\bibnamefont {Santos}},
  \bibinfo {author} {\bibfnamefont {C.}~\bibnamefont {Chamon}}, \ and\ \bibinfo
  {author} {\bibfnamefont {C.}~\bibnamefont {Mudry}},\ }\href@noop {}
  {\bibfield  {journal} {\bibinfo  {journal} {Phys. Rev. Lett.}\ }\textbf
  {\bibinfo {volume} {106}},\ \bibinfo {pages} {236804} (\bibinfo {year}
  {2011})}\BibitemShut {NoStop}%
\bibitem [{\citenamefont {Sheng}\ \emph {et~al.}(2011)\citenamefont {Sheng},
  \citenamefont {Gu}, \citenamefont {Sun},\ and\ \citenamefont
  {Sheng}}]{SGS11}%
  \BibitemOpen
  \bibfield  {author} {\bibinfo {author} {\bibfnamefont {D.}~\bibnamefont
  {Sheng}}, \bibinfo {author} {\bibfnamefont {Z.-C.}\ \bibnamefont {Gu}},
  \bibinfo {author} {\bibfnamefont {K.}~\bibnamefont {Sun}}, \ and\ \bibinfo
  {author} {\bibfnamefont {L.}~\bibnamefont {Sheng}},\ }\href@noop {}
  {\bibfield  {journal} {\bibinfo  {journal} {Nature Commun.}\ }\textbf
  {\bibinfo {volume} {2}},\ \bibinfo {pages} {389} (\bibinfo {year}
  {2011})}\BibitemShut {NoStop}%
\bibitem [{\citenamefont {Regnault}\ and\ \citenamefont
  {Bernevig}(2011)}]{RB11}%
  \BibitemOpen
  \bibfield  {author} {\bibinfo {author} {\bibfnamefont {N.}~\bibnamefont
  {Regnault}}\ and\ \bibinfo {author} {\bibfnamefont {B.~A.}\ \bibnamefont
  {Bernevig}},\ }\href@noop {} {\bibfield  {journal} {\bibinfo  {journal}
  {Phys. Rev. X}\ }\textbf {\bibinfo {volume} {1}},\ \bibinfo {pages} {021014}
  (\bibinfo {year} {2011})}\BibitemShut {NoStop}%
\bibitem [{\citenamefont {Tang}\ \emph {et~al.}(2011)\citenamefont {Tang},
  \citenamefont {Mei},\ and\ \citenamefont {Wen}}]{TMW11}%
  \BibitemOpen
  \bibfield  {author} {\bibinfo {author} {\bibfnamefont {E.}~\bibnamefont
  {Tang}}, \bibinfo {author} {\bibfnamefont {J.-W.}\ \bibnamefont {Mei}}, \
  and\ \bibinfo {author} {\bibfnamefont {X.-G.}\ \bibnamefont {Wen}},\
  }\href@noop {} {\bibfield  {journal} {\bibinfo  {journal} {Phys. Rev. Lett.}\
  }\textbf {\bibinfo {volume} {106}},\ \bibinfo {pages} {236802} (\bibinfo
  {year} {2011})}\BibitemShut {NoStop}%
\bibitem [{\citenamefont {Sun}\ \emph {et~al.}(2011)\citenamefont {Sun},
  \citenamefont {Gu}, \citenamefont {Katsura},\ and\ \citenamefont
  {Das~Sarma}}]{SGK11}%
  \BibitemOpen
  \bibfield  {author} {\bibinfo {author} {\bibfnamefont {K.}~\bibnamefont
  {Sun}}, \bibinfo {author} {\bibfnamefont {Z.-C.}\ \bibnamefont {Gu}},
  \bibinfo {author} {\bibfnamefont {H.}~\bibnamefont {Katsura}}, \ and\
  \bibinfo {author} {\bibfnamefont {S.}~\bibnamefont {Das~Sarma}},\ }\href@noop
  {} {\bibfield  {journal} {\bibinfo  {journal} {Phys. Rev. Lett.}\ }\textbf
  {\bibinfo {volume} {106}},\ \bibinfo {pages} {236803} (\bibinfo {year}
  {2011})}\BibitemShut {NoStop}%
\bibitem [{\citenamefont {Wu}\ \emph {et~al.}(2012)\citenamefont {Wu},
  \citenamefont {Bernevig},\ and\ \citenamefont {Regnault}}]{WBR12}%
  \BibitemOpen
  \bibfield  {author} {\bibinfo {author} {\bibfnamefont {Y.-L.}\ \bibnamefont
  {Wu}}, \bibinfo {author} {\bibfnamefont {B.~A.}\ \bibnamefont {Bernevig}}, \
  and\ \bibinfo {author} {\bibfnamefont {N.}~\bibnamefont {Regnault}},\
  }\href@noop {} {\bibfield  {journal} {\bibinfo  {journal} {Phys. Rev. B}\
  }\textbf {\bibinfo {volume} {85}},\ \bibinfo {pages} {075116} (\bibinfo
  {year} {2012})}\BibitemShut {NoStop}%
\bibitem [{\citenamefont {Bergholtz}\ and\ \citenamefont {Liu}(2013)}]{BL13}%
  \BibitemOpen
  \bibfield  {author} {\bibinfo {author} {\bibfnamefont {E.~J.}\ \bibnamefont
  {Bergholtz}}\ and\ \bibinfo {author} {\bibfnamefont {Z.}~\bibnamefont
  {Liu}},\ }\href {\doibase 10.1142/S021797921330017X} {\bibfield  {journal}
  {\bibinfo  {journal} {Int. J. Mod. Phys. B}\ }\textbf {\bibinfo {volume}
  {27}},\ \bibinfo {pages} {1330017} (\bibinfo {year} {2013})}\BibitemShut
  {NoStop}%
\bibitem [{\citenamefont {Parameswaran}\ \emph {et~al.}(2013)\citenamefont
  {Parameswaran}, \citenamefont {Roy},\ and\ \citenamefont {Sondhi}}]{PRS13}%
  \BibitemOpen
  \bibfield  {author} {\bibinfo {author} {\bibfnamefont {S.~A.}\ \bibnamefont
  {Parameswaran}}, \bibinfo {author} {\bibfnamefont {R.}~\bibnamefont {Roy}}, \
  and\ \bibinfo {author} {\bibfnamefont {S.~L.}\ \bibnamefont {Sondhi}},\
  }\href@noop {} {\  (\bibinfo {year} {2013})},\ \Eprint
  {http://arxiv.org/abs/arXiv:1302.6606} {arXiv:1302.6606} \BibitemShut
  {NoStop}%
\bibitem [{\citenamefont {Laughlin}(1983)}]{L83}%
  \BibitemOpen
  \bibfield  {author} {\bibinfo {author} {\bibfnamefont {R.~B.}\ \bibnamefont
  {Laughlin}},\ }\href {\doibase 10.1103/PhysRevLett.50.1395} {\bibfield
  {journal} {\bibinfo  {journal} {Phys. Rev. Lett.}\ }\textbf {\bibinfo
  {volume} {50}},\ \bibinfo {pages} {1395} (\bibinfo {year}
  {1983})}\BibitemShut {NoStop}%
\bibitem [{\citenamefont {Kol}\ and\ \citenamefont {Read}(1993)}]{KR93}%
  \BibitemOpen
  \bibfield  {author} {\bibinfo {author} {\bibfnamefont {A.}~\bibnamefont
  {Kol}}\ and\ \bibinfo {author} {\bibfnamefont {N.}~\bibnamefont {Read}},\
  }\href {\doibase 10.1103/PhysRevB.48.8890} {\bibfield  {journal} {\bibinfo
  {journal} {Phys. Rev. B}\ }\textbf {\bibinfo {volume} {48}},\ \bibinfo
  {pages} {8890} (\bibinfo {year} {1993})}\BibitemShut {NoStop}%
\bibitem [{\citenamefont {S\o{}rensen}\ \emph {et~al.}(2005)\citenamefont
  {S\o{}rensen}, \citenamefont {Demler},\ and\ \citenamefont {Lukin}}]{SAD05}%
  \BibitemOpen
  \bibfield  {author} {\bibinfo {author} {\bibfnamefont {A.~S.}\ \bibnamefont
  {S\o{}rensen}}, \bibinfo {author} {\bibfnamefont {E.}~\bibnamefont {Demler}},
  \ and\ \bibinfo {author} {\bibfnamefont {M.~D.}\ \bibnamefont {Lukin}},\
  }\href {\doibase 10.1103/PhysRevLett.94.086803} {\bibfield  {journal}
  {\bibinfo  {journal} {Phys. Rev. Lett.}\ }\textbf {\bibinfo {volume} {94}},\
  \bibinfo {pages} {086803} (\bibinfo {year} {2005})}\BibitemShut {NoStop}%
\bibitem [{\citenamefont {Palmer}\ and\ \citenamefont {Jaksch}(2006)}]{PJ06}%
  \BibitemOpen
  \bibfield  {author} {\bibinfo {author} {\bibfnamefont {R.~N.}\ \bibnamefont
  {Palmer}}\ and\ \bibinfo {author} {\bibfnamefont {D.}~\bibnamefont
  {Jaksch}},\ }\href {\doibase 10.1103/PhysRevLett.96.180407} {\bibfield
  {journal} {\bibinfo  {journal} {Phys. Rev. Lett.}\ }\textbf {\bibinfo
  {volume} {96}},\ \bibinfo {pages} {180407} (\bibinfo {year}
  {2006})}\BibitemShut {NoStop}%
\bibitem [{\citenamefont {Hafezi}\ \emph {et~al.}(2007)\citenamefont {Hafezi},
  \citenamefont {S\o{}rensen}, \citenamefont {Demler},\ and\ \citenamefont
  {Lukin}}]{HSD07}%
  \BibitemOpen
  \bibfield  {author} {\bibinfo {author} {\bibfnamefont {M.}~\bibnamefont
  {Hafezi}}, \bibinfo {author} {\bibfnamefont {A.~S.}\ \bibnamefont
  {S\o{}rensen}}, \bibinfo {author} {\bibfnamefont {E.}~\bibnamefont {Demler}},
  \ and\ \bibinfo {author} {\bibfnamefont {M.~D.}\ \bibnamefont {Lukin}},\
  }\href {\doibase 10.1103/PhysRevA.76.023613} {\bibfield  {journal} {\bibinfo
  {journal} {Phys. Rev. A}\ }\textbf {\bibinfo {volume} {76}},\ \bibinfo
  {pages} {023613} (\bibinfo {year} {2007})}\BibitemShut {NoStop}%
\bibitem [{\citenamefont {M\"oller}\ and\ \citenamefont {Cooper}(2009)}]{MC09}%
  \BibitemOpen
  \bibfield  {author} {\bibinfo {author} {\bibfnamefont {G.}~\bibnamefont
  {M\"oller}}\ and\ \bibinfo {author} {\bibfnamefont {N.~R.}\ \bibnamefont
  {Cooper}},\ }\href {\doibase 10.1103/PhysRevLett.103.105303} {\bibfield
  {journal} {\bibinfo  {journal} {Phys. Rev. Lett.}\ }\textbf {\bibinfo
  {volume} {103}},\ \bibinfo {pages} {105303} (\bibinfo {year}
  {2009})}\BibitemShut {NoStop}%
\bibitem [{\citenamefont {Sterdyniak}\ \emph {et~al.}(2012)\citenamefont
  {Sterdyniak}, \citenamefont {Regnault},\ and\ \citenamefont
  {M\"oller}}]{SRM12}%
  \BibitemOpen
  \bibfield  {author} {\bibinfo {author} {\bibfnamefont {A.}~\bibnamefont
  {Sterdyniak}}, \bibinfo {author} {\bibfnamefont {N.}~\bibnamefont
  {Regnault}}, \ and\ \bibinfo {author} {\bibfnamefont {G.}~\bibnamefont
  {M\"oller}},\ }\href {\doibase 10.1103/PhysRevB.86.165314} {\bibfield
  {journal} {\bibinfo  {journal} {Phys. Rev. B}\ }\textbf {\bibinfo {volume}
  {86}},\ \bibinfo {pages} {165314} (\bibinfo {year} {2012})}\BibitemShut
  {NoStop}%
\bibitem [{\citenamefont {Scaffidi}\ and\ \citenamefont {Simon}(2014)}]{SS14}%
  \BibitemOpen
  \bibfield  {author} {\bibinfo {author} {\bibfnamefont {T.}~\bibnamefont
  {Scaffidi}}\ and\ \bibinfo {author} {\bibfnamefont {S.~H.}\ \bibnamefont
  {Simon}},\ }\href@noop {} {\bibfield  {journal} {\bibinfo  {journal} {ArXiv
  e-prints}\ } (\bibinfo {year} {2014})},\ \Eprint
  {http://arxiv.org/abs/1407.1321} {arXiv:1407.1321 [cond-mat.str-el]}
  \BibitemShut {NoStop}%
\bibitem [{\citenamefont {Ghaemi}\ \emph {et~al.}(2012)\citenamefont {Ghaemi},
  \citenamefont {Cayssol}, \citenamefont {Sheng},\ and\ \citenamefont
  {Vishwanath}}]{GCS12}%
  \BibitemOpen
  \bibfield  {author} {\bibinfo {author} {\bibfnamefont {P.}~\bibnamefont
  {Ghaemi}}, \bibinfo {author} {\bibfnamefont {J.}~\bibnamefont {Cayssol}},
  \bibinfo {author} {\bibfnamefont {D.~N.}\ \bibnamefont {Sheng}}, \ and\
  \bibinfo {author} {\bibfnamefont {A.}~\bibnamefont {Vishwanath}},\ }\href
  {\doibase 10.1103/PhysRevLett.108.266801} {\bibfield  {journal} {\bibinfo
  {journal} {Phys. Rev. Lett.}\ }\textbf {\bibinfo {volume} {108}},\ \bibinfo
  {pages} {266801} (\bibinfo {year} {2012})}\BibitemShut {NoStop}%
\bibitem [{\citenamefont {Cooper}\ and\ \citenamefont {Dalibard}(2013)}]{CD13}%
  \BibitemOpen
  \bibfield  {author} {\bibinfo {author} {\bibfnamefont {N.~R.}\ \bibnamefont
  {Cooper}}\ and\ \bibinfo {author} {\bibfnamefont {J.}~\bibnamefont
  {Dalibard}},\ }\href {\doibase 10.1103/PhysRevLett.110.185301} {\bibfield
  {journal} {\bibinfo  {journal} {Phys. Rev. Lett.}\ }\textbf {\bibinfo
  {volume} {110}},\ \bibinfo {pages} {185301} (\bibinfo {year}
  {2013})}\BibitemShut {NoStop}%
\bibitem [{\citenamefont {Yao}\ \emph {et~al.}(2013)\citenamefont {Yao},
  \citenamefont {Gorshkov}, \citenamefont {Laumann}, \citenamefont {L\"auchli},
  \citenamefont {Ye},\ and\ \citenamefont {Lukin}}]{YGL13}%
  \BibitemOpen
  \bibfield  {author} {\bibinfo {author} {\bibfnamefont {N.~Y.}\ \bibnamefont
  {Yao}}, \bibinfo {author} {\bibfnamefont {A.~V.}\ \bibnamefont {Gorshkov}},
  \bibinfo {author} {\bibfnamefont {C.~R.}\ \bibnamefont {Laumann}}, \bibinfo
  {author} {\bibfnamefont {A.~M.}\ \bibnamefont {L\"auchli}}, \bibinfo {author}
  {\bibfnamefont {J.}~\bibnamefont {Ye}}, \ and\ \bibinfo {author}
  {\bibfnamefont {M.~D.}\ \bibnamefont {Lukin}},\ }\href {\doibase
  10.1103/PhysRevLett.110.185302} {\bibfield  {journal} {\bibinfo  {journal}
  {Phys. Rev. Lett.}\ }\textbf {\bibinfo {volume} {110}},\ \bibinfo {pages}
  {185302} (\bibinfo {year} {2013})}\BibitemShut {NoStop}%
\bibitem [{\citenamefont {Grushin}\ \emph {et~al.}(2014)\citenamefont
  {Grushin}, \citenamefont {G\'omez-Le\'on},\ and\ \citenamefont
  {Neupert}}]{GGN14}%
  \BibitemOpen
  \bibfield  {author} {\bibinfo {author} {\bibfnamefont {A.~G.}\ \bibnamefont
  {Grushin}}, \bibinfo {author} {\bibfnamefont {A.}~\bibnamefont
  {G\'omez-Le\'on}}, \ and\ \bibinfo {author} {\bibfnamefont {T.}~\bibnamefont
  {Neupert}},\ }\href {\doibase 10.1103/PhysRevLett.112.156801} {\bibfield
  {journal} {\bibinfo  {journal} {Phys. Rev. Lett.}\ }\textbf {\bibinfo
  {volume} {112}},\ \bibinfo {pages} {156801} (\bibinfo {year}
  {2014})}\BibitemShut {NoStop}%
\bibitem [{\citenamefont {Kourtis}\ \emph {et~al.}(2014)\citenamefont
  {Kourtis}, \citenamefont {Neupert}, \citenamefont {Chamon},\ and\
  \citenamefont {Mudry}}]{KNC14}%
  \BibitemOpen
  \bibfield  {author} {\bibinfo {author} {\bibfnamefont {S.}~\bibnamefont
  {Kourtis}}, \bibinfo {author} {\bibfnamefont {T.}~\bibnamefont {Neupert}},
  \bibinfo {author} {\bibfnamefont {C.}~\bibnamefont {Chamon}}, \ and\ \bibinfo
  {author} {\bibfnamefont {C.}~\bibnamefont {Mudry}},\ }\href {\doibase
  10.1103/PhysRevLett.112.126806} {\bibfield  {journal} {\bibinfo  {journal}
  {Phys. Rev. Lett.}\ }\textbf {\bibinfo {volume} {112}},\ \bibinfo {pages}
  {126806} (\bibinfo {year} {2014})}\BibitemShut {NoStop}%
\bibitem [{\citenamefont {Kourtis}\ \emph {et~al.}(2012)\citenamefont
  {Kourtis}, \citenamefont {Venderbos},\ and\ \citenamefont
  {Daghofer}}]{KVD12}%
  \BibitemOpen
  \bibfield  {author} {\bibinfo {author} {\bibfnamefont {S.}~\bibnamefont
  {Kourtis}}, \bibinfo {author} {\bibfnamefont {J.~W.}\ \bibnamefont
  {Venderbos}}, \ and\ \bibinfo {author} {\bibfnamefont {M.}~\bibnamefont
  {Daghofer}},\ }\href@noop {} {\bibfield  {journal} {\bibinfo  {journal}
  {Phys. Rev. B}\ }\textbf {\bibinfo {volume} {86}},\ \bibinfo {pages} {235118}
  (\bibinfo {year} {2012})}\BibitemShut {NoStop}%
\bibitem [{\citenamefont {Kourtis}\ and\ \citenamefont
  {Daghofer}(2013)}]{KD13}%
  \BibitemOpen
  \bibfield  {author} {\bibinfo {author} {\bibfnamefont {S.}~\bibnamefont
  {Kourtis}}\ and\ \bibinfo {author} {\bibfnamefont {M.}~\bibnamefont
  {Daghofer}},\ }\href@noop {} {\bibfield  {journal} {\bibinfo  {journal}
  {arXiv:1305.6948}\ } (\bibinfo {year} {2013})}\BibitemShut {NoStop}%
\bibitem [{\citenamefont {L\"auchli}\ \emph {et~al.}(2013)\citenamefont
  {L\"auchli}, \citenamefont {Liu}, \citenamefont {Bergholtz},\ and\
  \citenamefont {Moessner}}]{LLB12}%
  \BibitemOpen
  \bibfield  {author} {\bibinfo {author} {\bibfnamefont {A.~M.}\ \bibnamefont
  {L\"auchli}}, \bibinfo {author} {\bibfnamefont {Z.}~\bibnamefont {Liu}},
  \bibinfo {author} {\bibfnamefont {E.~J.}\ \bibnamefont {Bergholtz}}, \ and\
  \bibinfo {author} {\bibfnamefont {R.}~\bibnamefont {Moessner}},\ }\href
  {\doibase 10.1103/PhysRevLett.111.126802} {\bibfield  {journal} {\bibinfo
  {journal} {Phys. Rev. Lett.}\ }\textbf {\bibinfo {volume} {111}},\ \bibinfo
  {pages} {126802} (\bibinfo {year} {2013})}\BibitemShut {NoStop}%
\bibitem [{\citenamefont {Grushin}\ \emph {et~al.}(2012)\citenamefont
  {Grushin}, \citenamefont {Neupert}, \citenamefont {Chamon},\ and\
  \citenamefont {Mudry}}]{GNC12}%
  \BibitemOpen
  \bibfield  {author} {\bibinfo {author} {\bibfnamefont {A.~G.}\ \bibnamefont
  {Grushin}}, \bibinfo {author} {\bibfnamefont {T.}~\bibnamefont {Neupert}},
  \bibinfo {author} {\bibfnamefont {C.}~\bibnamefont {Chamon}}, \ and\ \bibinfo
  {author} {\bibfnamefont {C.}~\bibnamefont {Mudry}},\ }\href {\doibase
  10.1103/PhysRevB.86.205125} {\bibfield  {journal} {\bibinfo  {journal} {Phys.
  Rev. B}\ }\textbf {\bibinfo {volume} {86}},\ \bibinfo {pages} {205125}
  (\bibinfo {year} {2012})}\BibitemShut {NoStop}%
\bibitem [{\citenamefont {Zhu}\ \emph {et~al.}(2013)\citenamefont {Zhu},
  \citenamefont {Sheng},\ and\ \citenamefont {Haldane}}]{ZSH13}%
  \BibitemOpen
  \bibfield  {author} {\bibinfo {author} {\bibfnamefont {W.}~\bibnamefont
  {Zhu}}, \bibinfo {author} {\bibfnamefont {D.~N.}\ \bibnamefont {Sheng}}, \
  and\ \bibinfo {author} {\bibfnamefont {F.~D.~M.}\ \bibnamefont {Haldane}},\
  }\href {\doibase 10.1103/PhysRevB.88.035122} {\bibfield  {journal} {\bibinfo
  {journal} {Phys. Rev. B}\ }\textbf {\bibinfo {volume} {88}},\ \bibinfo
  {pages} {035122} (\bibinfo {year} {2013})}\BibitemShut {NoStop}%
\bibitem [{\citenamefont {Barkeshli}\ and\ \citenamefont {Qi}(2012)}]{BQ12}%
  \BibitemOpen
  \bibfield  {author} {\bibinfo {author} {\bibfnamefont {M.}~\bibnamefont
  {Barkeshli}}\ and\ \bibinfo {author} {\bibfnamefont {X.-L.}\ \bibnamefont
  {Qi}},\ }\href {\doibase 10.1103/PhysRevX.2.031013} {\bibfield  {journal}
  {\bibinfo  {journal} {Phys. Rev. X}\ }\textbf {\bibinfo {volume} {2}},\
  \bibinfo {pages} {031013} (\bibinfo {year} {2012})}\BibitemShut {NoStop}%
\bibitem [{\citenamefont {Liu}\ \emph {et~al.}(2012)\citenamefont {Liu},
  \citenamefont {Bergholtz}, \citenamefont {Fan},\ and\ \citenamefont
  {L\"auchli}}]{LBF12}%
  \BibitemOpen
  \bibfield  {author} {\bibinfo {author} {\bibfnamefont {Z.}~\bibnamefont
  {Liu}}, \bibinfo {author} {\bibfnamefont {E.~J.}\ \bibnamefont {Bergholtz}},
  \bibinfo {author} {\bibfnamefont {H.}~\bibnamefont {Fan}}, \ and\ \bibinfo
  {author} {\bibfnamefont {A.~M.}\ \bibnamefont {L\"auchli}},\ }\href {\doibase
  10.1103/PhysRevLett.109.186805} {\bibfield  {journal} {\bibinfo  {journal}
  {Phys. Rev. Lett.}\ }\textbf {\bibinfo {volume} {109}},\ \bibinfo {pages}
  {186805} (\bibinfo {year} {2012})}\BibitemShut {NoStop}%
\bibitem [{\citenamefont {Wang}\ \emph {et~al.}(2012)\citenamefont {Wang},
  \citenamefont {Yao}, \citenamefont {Gong},\ and\ \citenamefont
  {Sheng}}]{WHC12}%
  \BibitemOpen
  \bibfield  {author} {\bibinfo {author} {\bibfnamefont {Y.-F.}\ \bibnamefont
  {Wang}}, \bibinfo {author} {\bibfnamefont {H.}~\bibnamefont {Yao}}, \bibinfo
  {author} {\bibfnamefont {C.-D.}\ \bibnamefont {Gong}}, \ and\ \bibinfo
  {author} {\bibfnamefont {D.~N.}\ \bibnamefont {Sheng}},\ }\href {\doibase
  10.1103/PhysRevB.86.201101} {\bibfield  {journal} {\bibinfo  {journal} {Phys.
  Rev. B}\ }\textbf {\bibinfo {volume} {86}},\ \bibinfo {pages} {201101}
  (\bibinfo {year} {2012})}\BibitemShut {NoStop}%
\bibitem [{\citenamefont {Yang}\ \emph {et~al.}(2012)\citenamefont {Yang},
  \citenamefont {Gu}, \citenamefont {Sun},\ and\ \citenamefont
  {Das~Sarma}}]{YGS12}%
  \BibitemOpen
  \bibfield  {author} {\bibinfo {author} {\bibfnamefont {S.}~\bibnamefont
  {Yang}}, \bibinfo {author} {\bibfnamefont {Z.-C.}\ \bibnamefont {Gu}},
  \bibinfo {author} {\bibfnamefont {K.}~\bibnamefont {Sun}}, \ and\ \bibinfo
  {author} {\bibfnamefont {S.}~\bibnamefont {Das~Sarma}},\ }\href {\doibase
  10.1103/PhysRevB.86.241112} {\bibfield  {journal} {\bibinfo  {journal} {Phys.
  Rev. B}\ }\textbf {\bibinfo {volume} {86}},\ \bibinfo {pages} {241112}
  (\bibinfo {year} {2012})}\BibitemShut {NoStop}%
\bibitem [{\citenamefont {Wu}\ \emph {et~al.}(2013)\citenamefont {Wu},
  \citenamefont {Regnault},\ and\ \citenamefont {Bernevig}}]{WRB13a}%
  \BibitemOpen
  \bibfield  {author} {\bibinfo {author} {\bibfnamefont {Y.-L.}\ \bibnamefont
  {Wu}}, \bibinfo {author} {\bibfnamefont {N.}~\bibnamefont {Regnault}}, \ and\
  \bibinfo {author} {\bibfnamefont {B.~A.}\ \bibnamefont {Bernevig}},\ }\href
  {\doibase 10.1103/PhysRevLett.110.106802} {\bibfield  {journal} {\bibinfo
  {journal} {Phys. Rev. Lett.}\ }\textbf {\bibinfo {volume} {110}},\ \bibinfo
  {pages} {106802} (\bibinfo {year} {2013})}\BibitemShut {NoStop}%
\bibitem [{\citenamefont {Sterdyniak}\ \emph {et~al.}(2013)\citenamefont
  {Sterdyniak}, \citenamefont {Repellin}, \citenamefont {Bernevig},\ and\
  \citenamefont {Regnault}}]{SRB13}%
  \BibitemOpen
  \bibfield  {author} {\bibinfo {author} {\bibfnamefont {A.}~\bibnamefont
  {Sterdyniak}}, \bibinfo {author} {\bibfnamefont {C.}~\bibnamefont
  {Repellin}}, \bibinfo {author} {\bibfnamefont {B.~A.}\ \bibnamefont
  {Bernevig}}, \ and\ \bibinfo {author} {\bibfnamefont {N.}~\bibnamefont
  {Regnault}},\ }\href {\doibase 10.1103/PhysRevB.87.205137} {\bibfield
  {journal} {\bibinfo  {journal} {Phys. Rev. B}\ }\textbf {\bibinfo {volume}
  {87}},\ \bibinfo {pages} {205137} (\bibinfo {year} {2013})}\BibitemShut
  {NoStop}%
\bibitem [{\citenamefont {Wu}\ \emph {et~al.}(2014)\citenamefont {Wu},
  \citenamefont {Regnault},\ and\ \citenamefont {Bernevig}}]{WRB14}%
  \BibitemOpen
  \bibfield  {author} {\bibinfo {author} {\bibfnamefont {Y.-L.}\ \bibnamefont
  {Wu}}, \bibinfo {author} {\bibfnamefont {N.}~\bibnamefont {Regnault}}, \ and\
  \bibinfo {author} {\bibfnamefont {B.~A.}\ \bibnamefont {Bernevig}},\ }\href
  {\doibase 10.1103/PhysRevB.89.155113} {\bibfield  {journal} {\bibinfo
  {journal} {Phys. Rev. B}\ }\textbf {\bibinfo {volume} {89}},\ \bibinfo
  {pages} {155113} (\bibinfo {year} {2014})}\BibitemShut {NoStop}%
\bibitem [{\citenamefont {Haldane}(1988)}]{H88}%
  \BibitemOpen
  \bibfield  {author} {\bibinfo {author} {\bibfnamefont {F.~D.~M.}\
  \bibnamefont {Haldane}},\ }\href@noop {} {\bibfield  {journal} {\bibinfo
  {journal} {Phys. Rev. Lett.}\ }\textbf {\bibinfo {volume} {61}},\ \bibinfo
  {pages} {2015} (\bibinfo {year} {1988})}\BibitemShut {NoStop}%
\bibitem [{\citenamefont {McCulloch}(2008)}]{M08}%
  \BibitemOpen
  \bibfield  {author} {\bibinfo {author} {\bibfnamefont {I.~P.}\ \bibnamefont
  {McCulloch}},\ }\href@noop {} {\bibfield  {journal} {\bibinfo  {journal}
  {arXiv:0804.2509}\ } (\bibinfo {year} {2008})}\BibitemShut {NoStop}%
\bibitem [{\citenamefont {White}(1992)}]{W92}%
  \BibitemOpen
  \bibfield  {author} {\bibinfo {author} {\bibfnamefont {S.~R.}\ \bibnamefont
  {White}},\ }\href {\doibase 10.1103/PhysRevLett.69.2863} {\bibfield
  {journal} {\bibinfo  {journal} {Phys. Rev. Lett.}\ }\textbf {\bibinfo
  {volume} {69}},\ \bibinfo {pages} {2863} (\bibinfo {year}
  {1992})}\BibitemShut {NoStop}%
\bibitem [{\citenamefont {Kjall}\ \emph {et~al.}(2013)\citenamefont {Kjall},
  \citenamefont {Zaletel}, \citenamefont {Mong}, \citenamefont {Bardarson},\
  and\ \citenamefont {Pollmann}}]{KZM13}%
  \BibitemOpen
  \bibfield  {author} {\bibinfo {author} {\bibfnamefont {J.~A.}\ \bibnamefont
  {Kjall}}, \bibinfo {author} {\bibfnamefont {M.~P.}\ \bibnamefont {Zaletel}},
  \bibinfo {author} {\bibfnamefont {R.~S.~K.}\ \bibnamefont {Mong}}, \bibinfo
  {author} {\bibfnamefont {J.~H.}\ \bibnamefont {Bardarson}}, \ and\ \bibinfo
  {author} {\bibfnamefont {F.}~\bibnamefont {Pollmann}},\ }\href {\doibase
  10.1103/PhysRevB.87.235106} {\bibfield  {journal} {\bibinfo  {journal} {Phys.
  Rev. B}\ }\textbf {\bibinfo {volume} {87}},\ \bibinfo {pages} {235106}
  (\bibinfo {year} {2013})}\BibitemShut {NoStop}%
\bibitem [{\citenamefont {Kitaev}\ and\ \citenamefont {Preskill}(2006)}]{KP06}%
  \BibitemOpen
  \bibfield  {author} {\bibinfo {author} {\bibfnamefont {A.}~\bibnamefont
  {Kitaev}}\ and\ \bibinfo {author} {\bibfnamefont {J.}~\bibnamefont
  {Preskill}},\ }\href {\doibase 10.1103/PhysRevLett.96.110404} {\bibfield
  {journal} {\bibinfo  {journal} {Phys. Rev. Lett.}\ }\textbf {\bibinfo
  {volume} {96}},\ \bibinfo {pages} {110404} (\bibinfo {year}
  {2006})}\BibitemShut {NoStop}%
\bibitem [{\citenamefont {Levin}\ and\ \citenamefont {Wen}(2006)}]{LW06}%
  \BibitemOpen
  \bibfield  {author} {\bibinfo {author} {\bibfnamefont {M.}~\bibnamefont
  {Levin}}\ and\ \bibinfo {author} {\bibfnamefont {X.-G.}\ \bibnamefont
  {Wen}},\ }\href {\doibase 10.1103/PhysRevLett.96.110405} {\bibfield
  {journal} {\bibinfo  {journal} {Phys. Rev. Lett.}\ }\textbf {\bibinfo
  {volume} {96}},\ \bibinfo {pages} {110405} (\bibinfo {year}
  {2006})}\BibitemShut {NoStop}%
\bibitem [{\citenamefont {Jiang}\ \emph {et~al.}(2012)\citenamefont {Jiang},
  \citenamefont {Wang},\ and\ \citenamefont {Balents}}]{JWB12}%
  \BibitemOpen
  \bibfield  {author} {\bibinfo {author} {\bibfnamefont {H.-C.}\ \bibnamefont
  {Jiang}}, \bibinfo {author} {\bibfnamefont {Z.}~\bibnamefont {Wang}}, \ and\
  \bibinfo {author} {\bibfnamefont {L.}~\bibnamefont {Balents}},\ }\href
  {\doibase 10.1038/nphys2465} {\bibfield  {journal} {\bibinfo  {journal}
  {Nature Physics}\ }\textbf {\bibinfo {volume} {8}},\ \bibinfo {pages} {902}
  (\bibinfo {year} {2012})},\ \Eprint {http://arxiv.org/abs/1205.4289}
  {1205.4289} \BibitemShut {NoStop}%
\bibitem [{\citenamefont {Li}\ and\ \citenamefont {Haldane}(2008)}]{LH08}%
  \BibitemOpen
  \bibfield  {author} {\bibinfo {author} {\bibfnamefont {H.}~\bibnamefont
  {Li}}\ and\ \bibinfo {author} {\bibfnamefont {F.~D.~M.}\ \bibnamefont
  {Haldane}},\ }\href {\doibase 10.1103/PhysRevLett.101.010504} {\bibfield
  {journal} {\bibinfo  {journal} {Phys. Rev. Lett.}\ }\textbf {\bibinfo
  {volume} {101}},\ \bibinfo {pages} {010504} (\bibinfo {year}
  {2008})}\BibitemShut {NoStop}%
\bibitem [{\citenamefont {Tu}\ \emph {et~al.}(2013)\citenamefont {Tu},
  \citenamefont {Zhang},\ and\ \citenamefont {Qi}}]{HZQ13}%
  \BibitemOpen
  \bibfield  {author} {\bibinfo {author} {\bibfnamefont {H.-H.}\ \bibnamefont
  {Tu}}, \bibinfo {author} {\bibfnamefont {Y.}~\bibnamefont {Zhang}}, \ and\
  \bibinfo {author} {\bibfnamefont {X.-L.}\ \bibnamefont {Qi}},\ }\href
  {\doibase 10.1103/PhysRevB.88.195412} {\bibfield  {journal} {\bibinfo
  {journal} {Phys. Rev. B}\ }\textbf {\bibinfo {volume} {88}},\ \bibinfo
  {pages} {195412} (\bibinfo {year} {2013})}\BibitemShut {NoStop}%
\bibitem [{\citenamefont {Zaletel}\ \emph {et~al.}(2013)\citenamefont
  {Zaletel}, \citenamefont {Mong},\ and\ \citenamefont {Pollmann}}]{ZMP13}%
  \BibitemOpen
  \bibfield  {author} {\bibinfo {author} {\bibfnamefont {M.~P.}\ \bibnamefont
  {Zaletel}}, \bibinfo {author} {\bibfnamefont {R.~S.~K.}\ \bibnamefont
  {Mong}}, \ and\ \bibinfo {author} {\bibfnamefont {F.}~\bibnamefont
  {Pollmann}},\ }\href {\doibase 10.1103/PhysRevLett.110.236801} {\bibfield
  {journal} {\bibinfo  {journal} {Phys. Rev. Lett.}\ }\textbf {\bibinfo
  {volume} {110}},\ \bibinfo {pages} {236801} (\bibinfo {year}
  {2013})}\BibitemShut {NoStop}%
\bibitem [{\citenamefont {He}\ \emph {et~al.}(2014{\natexlab{a}})\citenamefont
  {He}, \citenamefont {Sheng},\ and\ \citenamefont {Chen}}]{HSC14}%
  \BibitemOpen
  \bibfield  {author} {\bibinfo {author} {\bibfnamefont {Y.-C.}\ \bibnamefont
  {He}}, \bibinfo {author} {\bibfnamefont {D.~N.}\ \bibnamefont {Sheng}}, \
  and\ \bibinfo {author} {\bibfnamefont {Y.}~\bibnamefont {Chen}},\ }\href
  {\doibase 10.1103/PhysRevB.89.075110} {\bibfield  {journal} {\bibinfo
  {journal} {Phys. Rev. B}\ }\textbf {\bibinfo {volume} {89}},\ \bibinfo
  {pages} {075110} (\bibinfo {year} {2014}{\natexlab{a}})}\BibitemShut
  {NoStop}%
\bibitem [{\citenamefont {He}\ \emph {et~al.}(2014{\natexlab{b}})\citenamefont
  {He}, \citenamefont {Sheng},\ and\ \citenamefont {Chen}}]{HSC14b}%
  \BibitemOpen
  \bibfield  {author} {\bibinfo {author} {\bibfnamefont {Y.-C.}\ \bibnamefont
  {He}}, \bibinfo {author} {\bibfnamefont {N.}~\bibnamefont {Sheng},
  \bibfnamefont {D.}}, \ and\ \bibinfo {author} {\bibfnamefont
  {Y.}~\bibnamefont {Chen}},\ }\href {\doibase 10.1103/PhysRevLett.112.137202}
  {\bibfield  {journal} {\bibinfo  {journal} {Phys. Rev. Lett.}\ }\textbf
  {\bibinfo {volume} {112}},\ \bibinfo {pages} {137202} (\bibinfo {year}
  {2014}{\natexlab{b}})}\BibitemShut {NoStop}%
\bibitem [{\citenamefont {Cincio}\ and\ \citenamefont {Vidal}(2013)}]{CV13}%
  \BibitemOpen
  \bibfield  {author} {\bibinfo {author} {\bibfnamefont {L.}~\bibnamefont
  {Cincio}}\ and\ \bibinfo {author} {\bibfnamefont {G.}~\bibnamefont {Vidal}},\
  }\href {\doibase 10.1103/PhysRevLett.110.067208} {\bibfield  {journal}
  {\bibinfo  {journal} {Phys. Rev. Lett.}\ }\textbf {\bibinfo {volume} {110}},\
  \bibinfo {pages} {067208} (\bibinfo {year} {2013})}\BibitemShut {NoStop}%
\bibitem [{\citenamefont {Liu}\ \emph {et~al.}(2013)\citenamefont {Liu},
  \citenamefont {Kovrizhin},\ and\ \citenamefont {Bergholtz}}]{ZKB13}%
  \BibitemOpen
  \bibfield  {author} {\bibinfo {author} {\bibfnamefont {Z.}~\bibnamefont
  {Liu}}, \bibinfo {author} {\bibfnamefont {D.~L.}\ \bibnamefont {Kovrizhin}},
  \ and\ \bibinfo {author} {\bibfnamefont {E.~J.}\ \bibnamefont {Bergholtz}},\
  }\href {\doibase 10.1103/PhysRevB.88.081106} {\bibfield  {journal} {\bibinfo
  {journal} {Phys. Rev. B}\ }\textbf {\bibinfo {volume} {88}},\ \bibinfo
  {pages} {081106} (\bibinfo {year} {2013})}\BibitemShut {NoStop}%
\bibitem [{\citenamefont {Niu}\ \emph {et~al.}(1985)\citenamefont {Niu},
  \citenamefont {Thouless},\ and\ \citenamefont {Wu}}]{NTDW85}%
  \BibitemOpen
  \bibfield  {author} {\bibinfo {author} {\bibfnamefont {Q.}~\bibnamefont
  {Niu}}, \bibinfo {author} {\bibfnamefont {D.~J.}\ \bibnamefont {Thouless}}, \
  and\ \bibinfo {author} {\bibfnamefont {Y.-S.}\ \bibnamefont {Wu}},\
  }\href@noop {} {\bibfield  {journal} {\bibinfo  {journal} {Phys. Rev. B}\
  }\textbf {\bibinfo {volume} {31}},\ \bibinfo {pages} {3372} (\bibinfo {year}
  {1985})}\BibitemShut {NoStop}%
\bibitem [{\citenamefont {Zaletel}\ \emph {et~al.}(2014)\citenamefont
  {Zaletel}, \citenamefont {Mong},\ and\ \citenamefont {Pollmann}}]{ZMP14}%
  \BibitemOpen
  \bibfield  {author} {\bibinfo {author} {\bibfnamefont {M.~P.}\ \bibnamefont
  {Zaletel}}, \bibinfo {author} {\bibfnamefont {R.~S.~K.}\ \bibnamefont
  {Mong}}, \ and\ \bibinfo {author} {\bibfnamefont {F.}~\bibnamefont
  {Pollmann}},\ }\href@noop {} {\bibfield  {journal} {\bibinfo  {journal}
  {arXiv:1405.6028}\ } (\bibinfo {year} {2014})}\BibitemShut {NoStop}%
\bibitem [{Note1()}]{Note1}%
  \BibitemOpen
  \bibinfo {note} {The benchmarking and success of the iDMRG algorithm in
  describing trivial and CI phases at half-filling will be reported elsewhere
  \cite {MGP14}.}\BibitemShut {Stop}%
\bibitem [{\citenamefont {Alexandradinata}\ \emph {et~al.}(2011)\citenamefont
  {Alexandradinata}, \citenamefont {Hughes},\ and\ \citenamefont
  {Bernevig}}]{AHB11}%
  \BibitemOpen
  \bibfield  {author} {\bibinfo {author} {\bibfnamefont {A.}~\bibnamefont
  {Alexandradinata}}, \bibinfo {author} {\bibfnamefont {T.~L.}\ \bibnamefont
  {Hughes}}, \ and\ \bibinfo {author} {\bibfnamefont {B.~A.}\ \bibnamefont
  {Bernevig}},\ }\href {\doibase 10.1103/PhysRevB.84.195103} {\bibfield
  {journal} {\bibinfo  {journal} {Phys. Rev. B}\ }\textbf {\bibinfo {volume}
  {84}},\ \bibinfo {pages} {195103} (\bibinfo {year} {2011})}\BibitemShut
  {NoStop}%
\bibitem [{\citenamefont {Motruk}\ \emph {et~al.}(2014)\citenamefont {Motruk},
  \citenamefont {Grushin},\ and\ \citenamefont {Pollmann}}]{MGP14}%
  \BibitemOpen
  \bibfield  {author} {\bibinfo {author} {\bibfnamefont {J.}~\bibnamefont
  {Motruk}}, \bibinfo {author} {\bibfnamefont {A.~G.}\ \bibnamefont {Grushin}},
  \ and\ \bibinfo {author} {\bibfnamefont {F.}~\bibnamefont {Pollmann}},\
  }\href@noop {} {\bibfield  {journal} {\bibinfo  {journal} {In preparation}\ }
  (\bibinfo {year} {2014})}\BibitemShut {NoStop}%
\bibitem [{Note2()}]{Note2}%
  \BibitemOpen
  \bibinfo {note} {For $S(L)$ in Fig.~\ref {fig: FCI} we include only those
  sizes that satisfy $L=6m$ with $m\in \protect \mathbb {Z}$. Other sizes have
  a fractional charge per unit length on the cylinder in the Tao-Thouless limit
  and spontaneously break the symmetry into a charge density wave order which
  smears out as $L$ increases. The finite corrections for $S(L)$ when $L\not
  =6m$ are thus expected to be more severe than for $L=6m$ and therefore we
  exclude them.}\BibitemShut {Stop}%
\bibitem [{\citenamefont {Pollmann}\ and\ \citenamefont {Turner}(2012)}]{PT12}%
  \BibitemOpen
  \bibfield  {author} {\bibinfo {author} {\bibfnamefont {F.}~\bibnamefont
  {Pollmann}}\ and\ \bibinfo {author} {\bibfnamefont {A.~M.}\ \bibnamefont
  {Turner}},\ }\href {\doibase 10.1103/PhysRevB.86.125441} {\bibfield
  {journal} {\bibinfo  {journal} {Phys. Rev. B}\ }\textbf {\bibinfo {volume}
  {86}},\ \bibinfo {pages} {125441} (\bibinfo {year} {2012})}\BibitemShut
  {NoStop}%
\bibitem [{\citenamefont {Tagliacozzo}\ \emph {et~al.}(2008)\citenamefont
  {Tagliacozzo}, \citenamefont {de~Oliveira}, \citenamefont {Iblisdir},\ and\
  \citenamefont {Latorre}}]{Tagliacozzo-2008}%
  \BibitemOpen
  \bibfield  {author} {\bibinfo {author} {\bibfnamefont {L.}~\bibnamefont
  {Tagliacozzo}}, \bibinfo {author} {\bibfnamefont {T.~R.}\ \bibnamefont
  {de~Oliveira}}, \bibinfo {author} {\bibfnamefont {S.}~\bibnamefont
  {Iblisdir}}, \ and\ \bibinfo {author} {\bibfnamefont {J.~I.}\ \bibnamefont
  {Latorre}},\ }\href {\doibase 10.1103/PhysRevB.78.024410} {\bibfield
  {journal} {\bibinfo  {journal} {Phys. Rev. B}\ }\textbf {\bibinfo {volume}
  {78}},\ \bibinfo {pages} {024410} (\bibinfo {year} {2008})}\BibitemShut
  {NoStop}%
\bibitem [{\citenamefont {Pollmann}\ \emph {et~al.}(2009)\citenamefont
  {Pollmann}, \citenamefont {Mukerjee}, \citenamefont {Turner},\ and\
  \citenamefont {Moore}}]{Pollmann-2009}%
  \BibitemOpen
  \bibfield  {author} {\bibinfo {author} {\bibfnamefont {F.}~\bibnamefont
  {Pollmann}}, \bibinfo {author} {\bibfnamefont {S.}~\bibnamefont {Mukerjee}},
  \bibinfo {author} {\bibfnamefont {A.~M.}\ \bibnamefont {Turner}}, \ and\
  \bibinfo {author} {\bibfnamefont {J.~E.}\ \bibnamefont {Moore}},\ }\href
  {\doibase 10.1103/PhysRevLett.102.255701} {\bibfield  {journal} {\bibinfo
  {journal} {Phys. Rev. Lett.}\ }\textbf {\bibinfo {volume} {102}},\ \bibinfo
  {pages} {255701} (\bibinfo {year} {2009})}\BibitemShut {NoStop}%
\bibitem [{\citenamefont {Shankar}(1994)}]{S94}%
  \BibitemOpen
  \bibfield  {author} {\bibinfo {author} {\bibfnamefont {R.}~\bibnamefont
  {Shankar}},\ }\href {\doibase 10.1103/RevModPhys.66.129} {\bibfield
  {journal} {\bibinfo  {journal} {Rev. Mod. Phys.}\ }\textbf {\bibinfo {volume}
  {66}},\ \bibinfo {pages} {129} (\bibinfo {year} {1994})}\BibitemShut
  {NoStop}%
\bibitem [{\citenamefont {Zaletel}(2013)}]{Z14}%
  \BibitemOpen
  \bibfield  {author} {\bibinfo {author} {\bibfnamefont {M.~P.}\ \bibnamefont
  {Zaletel}},\ }\href@noop {} {\bibfield  {journal} {\bibinfo  {journal} {ArXiv
  e-prints}\ } (\bibinfo {year} {2013})},\ \Eprint
  {http://arxiv.org/abs/1309.7387} {arXiv:1309.7387 [cond-mat.str-el]}
  \BibitemShut {NoStop}%
\bibitem [{\citenamefont {Wen}(1990)}]{W90}%
  \BibitemOpen
  \bibfield  {author} {\bibinfo {author} {\bibfnamefont {X.~G.}\ \bibnamefont
  {Wen}},\ }\href@noop {} {\bibfield  {journal} {\bibinfo  {journal} {Int. J.
  Mod. Phys.}\ }\textbf {\bibinfo {volume} {B4}},\ \bibinfo {pages} {239}
  (\bibinfo {year} {1990})}\BibitemShut {NoStop}%
\end{thebibliography}
\appendix
\section{\label{app:1}Computing the quasiparticle charge and topological spin}
In this section we review how to extract the quasiparticle charge and topological spin from iDMRG.
For a detailed discussion we refer the reader to Refs. \onlinecite{ZMP13,HZQ13,Z14} which we follow closely. \\
Consider the infinite cylinder of Fig.~1(a) of the main text. Topologically, the cylinder is equivalent to
a sphere with two holes, each of which carries a quasiparticle $a$ for a given topologically non-trivial FCI groundstate
$\left.|GS_{a}\right\rangle$. A rotation of the cylinder along the finite direction $y$ rotates the two quasiparticles
simultaneously in opposite directions and thus the overlap between the final rotated state and the initial state
is one, since such a rotation is a global symmetry.\\
However, suppose we cut the cylinder into two semi-infinite halves $L$ and $R$.
The translation operator which rotates the cylinder factors into its action for each halve $\hat{T} = T_L \otimes T_R$.
By `adiabatically' rotating the right halve $R$ through $L_y$ unit cells, the corresponding quasiparticle will acquire a Berry phase $\theta_a = e^{2 \pi i h_{a}}$ where $\theta_{a} $ is known as the topological spin \cite{KP06,W90}.
To accomplish an adiabatic rotation on a lattice, it was argued in Ref. \onlinecite{HZQ13} that  we can compute a discrete Berry connection for rotating $R$ by one unit cell, $\left\langle GS_{a}\right| T_{R}\left|GS_{a}\right\rangle$.
The discrete Berry phase was argued to take the form
\begin{eqnarray}\nonumber
\lambda_{a}&\equiv& \left\langle GS_{a}\right|T_{R}\left|GS_{a}\right\rangle^{L_y}\\
 &=& \mathrm{exp}\left[2\pi i\left(h_{a}-\dfrac{c_-}{24}\right)-\alpha L_{y}\right]
\end{eqnarray}
$h_a$ is the desired topological spin, $c_-$ is the chiral central charge of the edge CFT, and $\alpha\in\mathbb{C}$ is generically a non-universal constant which is \emph{independent} of $a$.
From this expression it is therefore apparent that it is possible to extract $h_{a}$ if one has access to the different groundstates labelled by $a$.
\begin{equation}
\mathrm{arg}\left[\dfrac{\lambda_{a}}{\lambda_{a'}}\right]=2\pi\left(h_{a}-h_{a'}\right),
\end{equation}
where the $\mathrm{arg}$ function gives the argument of a given complex number.
In iDMRG a given topologically degenerate groundstate $\left.|GS_{a}\right\rangle$ evolves into a different
$\left.|GS_{a'}\right\rangle$ when a flux quantum is inserted. The number of topologically distinct $\left.|GS_{a}\right\rangle$
sectors is obtained by observing when a unit charge has been pumped through the boundary between $L$ and $R$, or
equivalently, when the entanglement spectrum has mapped into the original zero flux spectrum with a unit shift in the charge quantum number $q$ discussed in the main text.\\
With each inserted flux quantum we pump a quasiparticle through the boundary we effectively measure the difference between two given quasiparticle charges. Therefore, the quasiparticle charge $q_{a}$ is directly given by $\left\langle Q^{L} \right\rangle$ mod 1~\cite{ZMP13}. \\

For the $1/3$-Laughlin state quasiparticles, relevant to
our work we label the anyons with an integer $a=\left\lbrace -1,0,1 \right\rbrace$. The quasiparticle charge
and topological spin are given by $q_{a}=(a+1/2)/3$ and $h_{a}=(a+1/2)^2/6$ mod 1 respectively.
A technical point in order here is that the $1/2$ factor takes into account fermionic periodic boundary conditions.
Note that the $q_{a}-q_{a'}=0,\pm1/3$ mod 1 and that $|h_{a}-h_{a'}|=0,1/3$ mod 1.\\
From the above we can now read off the quasiparticle charge and topological spin
for the $1/3$-Laughlin FCI state from the iDMRG GS. In particular for $L=2L_{y}=12$ and $\chi=800$ we obtain that
the quasiparticle charges are:
\begin{eqnarray}
\nonumber
\left.\left\langle Q^{L}(\Phi_{y}) \right\rangle \right|_{\Phi_{y}=0} &=& 0.503\sim\left.\frac{(a+\frac{1}{2})}{3}\right|_{a=1}=\frac{1}{2}\\
\left.\left\langle Q^{L}(\Phi_{y}) \right\rangle \right|_{\Phi_{y}=2\pi} &=& 0.154\sim\left.\frac{(a+\frac{1}{2})}{3}\right|_{a=0}=\frac{1}{6}\\
\nonumber
\left.\left\langle Q^{L}(\Phi_{y}) \right\rangle \right|_{\Phi_{y}=4\pi} &=& -0.154\sim\left.\frac{(a+\frac{1}{2})}{3}\right|_{a=-1}=-\frac{1}{6}
\end{eqnarray}
which agree with the expected value.
It is simple to check also that indeed $q_{a}-q_{a'}\simeq0,\pm 1/3$ mod 1.
For the topological spin on the other hand we obtain (with the same parameters) :
\begin{equation}
\dfrac{1}{2\pi}\mathrm{arg}\left[\dfrac{\lambda_{\Phi_{y}=2\pi}}{\lambda_{\Phi_{y}=0}}\right]=0.36
\end{equation}
This should be compared to $h_{1}-h_{0}=1/3$.
\end{document}